\documentclass{emulateapj}

\begin{document}

\shortauthors{Loutrel et al.}
\shorttitle{Brown Dwarf Companion from WISE}

\title{Discovery of a Companion at the L/T Transition 
with the Wide-field Infrared Survey Explorer}

\author{
N. P. Loutrel\altaffilmark{1},
K. L. Luhman\altaffilmark{1,2},
P. J. Lowrance\altaffilmark{3}, and 
J. J. Bochanski\altaffilmark{1}}

\altaffiltext{1}{Department of Astronomy and Astrophysics,
The Pennsylvania State University, University Park, PA 16802, USA;
loutrel@astro.psu.edu}
\altaffiltext{2}{Center for Exoplanets and Habitable Worlds, The 
Pennsylvania State University, University Park, PA 16802, USA}
\altaffiltext{3}{Infrared Processing and Analysis Center, California
Institute of Technology 100-22, Pasadena, CA 91125, USA}

\begin{abstract}

We report the discovery of a substellar companion to the nearby solar-type
star HD~46588 (F7V, 17.9~pc, $\tau\sim3$~Gyr).  
HD~46588~B was found through a survey for common proper motion companions
to nearby stars using data from the Wide-field Infrared Survey Explorer
and the Two-Micron All-Sky Survey.  It has an angular separation of
$79.2\arcsec$ from its primary, which corresponds to a projected physical
separation of 1420~AU. We have measured a spectral type of L9 for this
object based on near-infrared spectroscopy performed with TripleSpec at
Palomar Observatory. We estimate a mass of $0.064^{+0.008}_{-0.019}$~$M_\odot$
from a comparison of its luminosity to the values predicted by theoretical
evolutionary models for the age of the primary. Because of its companionship
to a well-studied star, HD~46588~B is one of the few known brown dwarfs
at the L/T transition for which both age and distance estimates are
available. Thus, it offers new constraints on the properties of brown
dwarfs during this brief evolutionary phase.
The discovery of HD~46588~B also illustrates the value of the Wide-field
Infrared Survey Explorer for identifying brown dwarfs in the solar
neighborhood via their proper motions.

\end{abstract}


\section{Introduction}

L and T dwarf companions to nearby stars are valuable for studying the
structure and evolution of brown dwarfs as well as their origin.
Because the ages and metallicities of these companions often can be determined
via their primaries, they can be used to calibrate
atmospheric and evolutionary models \citep{kir05,cus08,hel08,bur10b}. 
Meanwhile, theories for the formation of brown dwarfs can be tested with
measurements of the fraction of stars that harbor wide low-mass companions
and the frequency of tight binaries among these companions
\citep{rei01,bat03,sta09}.

Multiple strategies have been employed in identifying cool companions
to stars in the solar neighborhood. One approach is to define a
sample of stars and search for companions to them
\citep{bec88,nak95,pot02,low05,bil06,met06,mug06,luh07,luh11}.
Alternatively, one can begin with a sample of known brown dwarfs
and check whether they are companions to stars
\citep{bur00,kir01,wil01,sch03,burn09,fah10,fah11,gol10,sch10,zha10,day11}.
Late-type companions also can be uncovered through a search for 
pairs of comoving objects across a large area of sky \citep{rad08}. 
Proper motion measurements are essential in all of these strategies,
either for the initial identification of candidate companions
or for confirming candidates that were selected through photometry.
These proper motions can be measured through dedicated imaging or
archival data from wide-field surveys like the 
Two Micron All-Sky Survey \citep[2MASS,][]{skr06},
the Sloan Digital Sky Survey \cite[SDSS,][]{yor00},
and the United Kingdom Infrared Telescope Infrared Deep Sky Survey
\citep[UKIDSS,][]{law07}.
A new source of astrometry has recently become available via
the Wide-field Infrared Survey Explorer \citep[WISE,][]{wri10}.
This mission has performed the first all-sky survey at infrared (IR)
wavelengths since 2MASS, thus providing an opportunity to measure proper
motions for 2MASS sources across the entire sky.
Indeed, proper motions incorporating WISE data have already been exploited for
searching for free-floating L and T dwarfs \citep{sch11,giz11}.

We have begun a search for faint common proper motion companions to nearby
stars using astrometry from 2MASS and WISE.
In this paper, we report the discovery of a substellar companion to 
the solar-type star HD~46588. We describe the methods that
uncovered this object as a possible companion (\S~\ref{sec:search})
and present spectroscopy that confirms its cool nature (\S~\ref{sec:spec}).
We then examine the physical properties of this new companion
(\S~\ref{sec:prop}) and discuss its relevance to studies of 
L and T dwarfs (\S~\ref{sec:disc}).

\section{2MASS-WISE Proper Motion Survey}
\label{sec:search}

Between 1997 and 2001, the 2MASS telescopes collected images of the entire sky
in broad-band filters centered at 1.25, 1.65, and 2.16~\micron\ \citep[$J$,
$H$, $K_s$,][]{skr06}. In 2010, the WISE satellite mapped the sky 
at 3.4, 4.6, 12, and 22~\micron\ \citep[W1--W4,][]{wri10}.
The typical magnitudes that correspond to SNR=5 for 2MASS and WISE are
$J=16.5$, $H=15.8$, $K_s=15.0$, W1=16.5, W2=15.5, W3=11.2, and W4=7.9
in unconfused areas of sky and (for WISE) in the ecliptic. WISE achieves
better sensitivity at higher ecliptic latitudes because of greater depth
of coverage.
Given these photometric limits and the near- and mid-IR colors of L and T
dwarfs \citep{leg10a}, a survey that requires detections in
both 2MASS and WISE will be limited by the sensitivity of 2MASS at $J$.

The astrometric precision of 2MASS is slightly less than $0\farcs1$ for
SNR$>$10 and approaches $0.2\arcsec$ at the detection limit \citep{skr06}.
For WISE, the precision is $0.2\arcsec$ for brighter sources (W1$<13$) and
approaches $1\arcsec$ in declination for some of the fainter
objects because of an error in the preliminary data processing.
Given the elapsed time of $\sim10$ years between 2MASS and WISE, 
proper motions measured from their astrometry should have accuracies
of $\lesssim10$\% at brighter magnitudes for $\mu\gtrsim0.2\arcsec$~yr$^{-1}$.
Therefore, we have considered for
our survey all known main sequence stars, white dwarfs, and brown dwarfs
in the solar neighborhood with proper motions exceeding this threshold
\citep[e.g.,][]{sub05,lep05,lep08,fah09}.

To search for possible late-type companions to a given primary, we
begin by retrieving lists of coordinates and photometry for all objects
within a radius of $1\arcdeg$ that appear in the 2MASS Point Source
Catalog and the WISE Preliminary Release Source Catalog. We match the
two lists to identify pairs of 2MASS and WISE sources whose astrometric offsets
correspond to the expected motion of the primary based on its known proper
motion. To be considered a candidate companion, we also require that an
object's colors and magnitudes (or limits) from 2MASS and WISE are consistent
with those of M, L, and T dwarfs at the distance of the primary.
To characterize the colors of the latter, we retrieved photometry from
the 2MASS and WISE catalogs for late-type dwarfs that have known distances
\citep[][references therein]{leg10a}. This process typically produces 
$\lesssim2$ candidate companions for most primaries. 
Stars that have low galactic latitudes or small proper motions can have
dozens of candidates. We then visually inspect the 2MASS and
WISE images to identify and reject false candidates, which often consist of
one component of a resolved pair of 2MASS sources matched to an unresolved
blend of the pair in WISE.

\begin{figure}
\epsscale{1.0}
\plotone{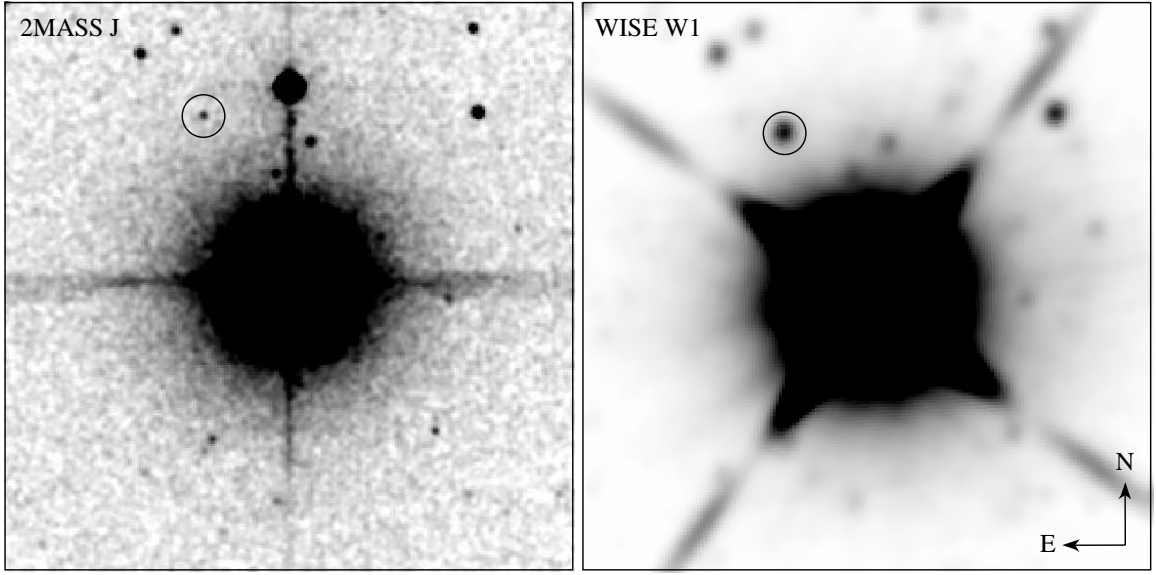}
\caption{
2MASS and WISE images of HD~46588~A and B.  The size of each image is 
$4\arcmin\times4\arcmin$.
}
\label{fig:image}
\end{figure}

\begin{figure}
\epsscale{1.0}
\plotone{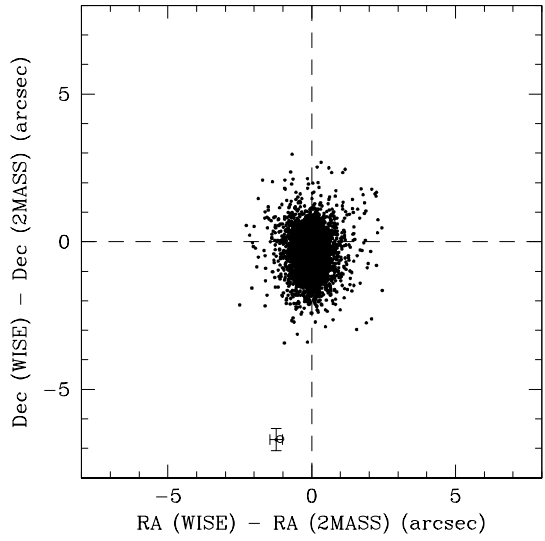}
\caption{
Differences in coordinates of sources within $1\arcdeg$ of HD~46588
between 2MASS and WISE (points). The motion of one of these sources
(1~$\sigma$ error bars) is consistent with the expected motion of
HD~46588 (circle). The vertical elongation in the distribution of offsets
is a reflection of the systematically larger uncertainties in the
preliminary measurements of declination for faint sources from WISE
(\S~\ref{sec:search}).
}
\label{fig:pm}
\end{figure}

Through our analysis of the 2MASS and WISE data for nearby stars,
we have identified a new companion to the F7V star HD~46588.
The companion is designated as 2MASS J06462756+7935045 and
WISEP J064627.10+793457.8, and we refer to it as HD~46588~B.
It has a projected separation of $\sim80\arcsec$, corresponding to 1420~AU
at the distance of the primary \citep[17.9~pc,][]{van07}.
We show 2MASS and WISE images of the pair in Figure~\ref{fig:image}.
The agreement between the motions of HD~46588~A and B is
demonstrated in Figure~\ref{fig:pm}, where we plot the differences in
equatorial coordinates between 2MASS and WISE for all stars within
a radius of $1\arcdeg$ and with W2 uncertainties less than 0.1~mag.
For stars that have W2 fluxes within $\pm0.5$~mag of HD~46588~B, the
standard deviations of the differences in right ascension and declination
are $0\farcs21$ and $0\farcs37$, respectively.
The colors and magnitudes of HD~46588~B from 2MASS and WISE
are consistent with those of a cool dwarf at the distance of the primary,
as illustrated in the color-magnitude diagrams in Figure~\ref{fig:cmd}.
The colors of HD~46588~B suggest that it is near the boundary between L and
T dwarfs.

\section{Spectroscopy}
\label{sec:spec}

To confirm its cool nature, we obtained a near-IR spectrum
of HD~46588~B with the TripleSpec spectrograph \citep{her08} on the
Palomar 5-m Hale telescope on the night of 2011 May 12.
TripleSpec produces spectra that extend from 1--2.4~\micron\ and have a
resolution of 2700.
We collected eight 5~min exposures of HD~46588~B in an ABBA dither
pattern along the slit ($1\arcsec\times30\arcsec$).
We also observed HD~45560 (A1V) to provide telluric correction.
The slit was aligned to the parallactic angle for these observations.
We reduced the data with a version of the Spextool package \citep{cus04}
that has been modified for use with data from TripleSpec (M. Cushing,
private communication). This software corrects for telluric absorption
with the methods described by \citet{vac03}.

\begin{figure}
\epsscale{1.0}
\plotone{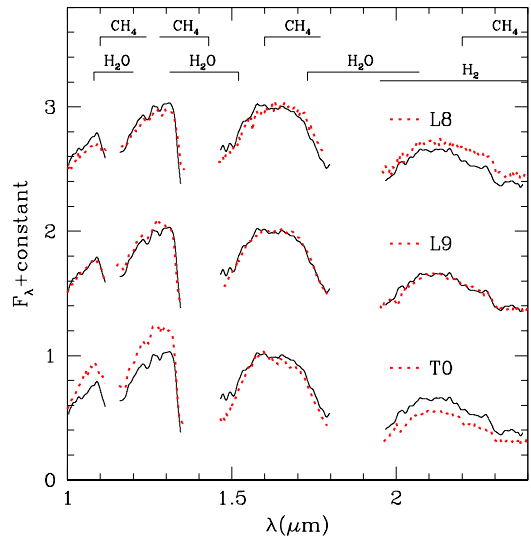}
\caption{
Spectrum of HD~46588~B (solid line) compared to data for the
near-IR standards 2MASSW J1632291+190441 \citep[L8,][]{bur04},
DENIS-P J0255-4700 \citep[L9,][]{bur06a}, and
SDSS J120747.17+024424.8 \citep[T0,][]{loo07}.
We classify HD~46588~B as L9 based on this comparison.
Each spectrum is normalized at its maximum flux in the $H$ band.
}
\label{fig:spec}
\end{figure}

We present the reduced spectrum of HD~46588~B in Figure~\ref{fig:spec}.
It exhibits strong absorption bands from H$_2$O, which confirms that it
is a cool object. To measure its spectral type, we compared the spectrum
of HD~46588~B to data from the SpeX Prism Spectral
Libraries\footnote{http://pono.ucsd.edu/$\sim$adam/browndwarfs/spexpris}
for L and T dwarf standards \citep{bur06a}.
Although the metallicity of HD~46588 appears to be subsolar 
\citep[\lbrack Fe/H\rbrack =$-$0.21,][]{cas11}, it is only $\sim0.1$~dex below the mean
value for stars in the solar neighborhood \citep{nor04,hol07}.
Thus, normal dwarf standards should be suitable for the classification
of HD~46588~B.
For comparison to the standards, the spectrum of HD~46588~B was smoothed
to the same resolution as the SpeX data (R$\sim$100). We find that HD~46588~B
closely resembles standards with a near-IR spectral type of L9, as
illustrated in Figure~\ref{fig:spec}. This classification is consistent
with that expected based on its photometry (Figure~\ref{fig:cmd}).

\section{Physical Properties}
\label{sec:prop}

\begin{figure}
\epsscale{1.0}
\plotone{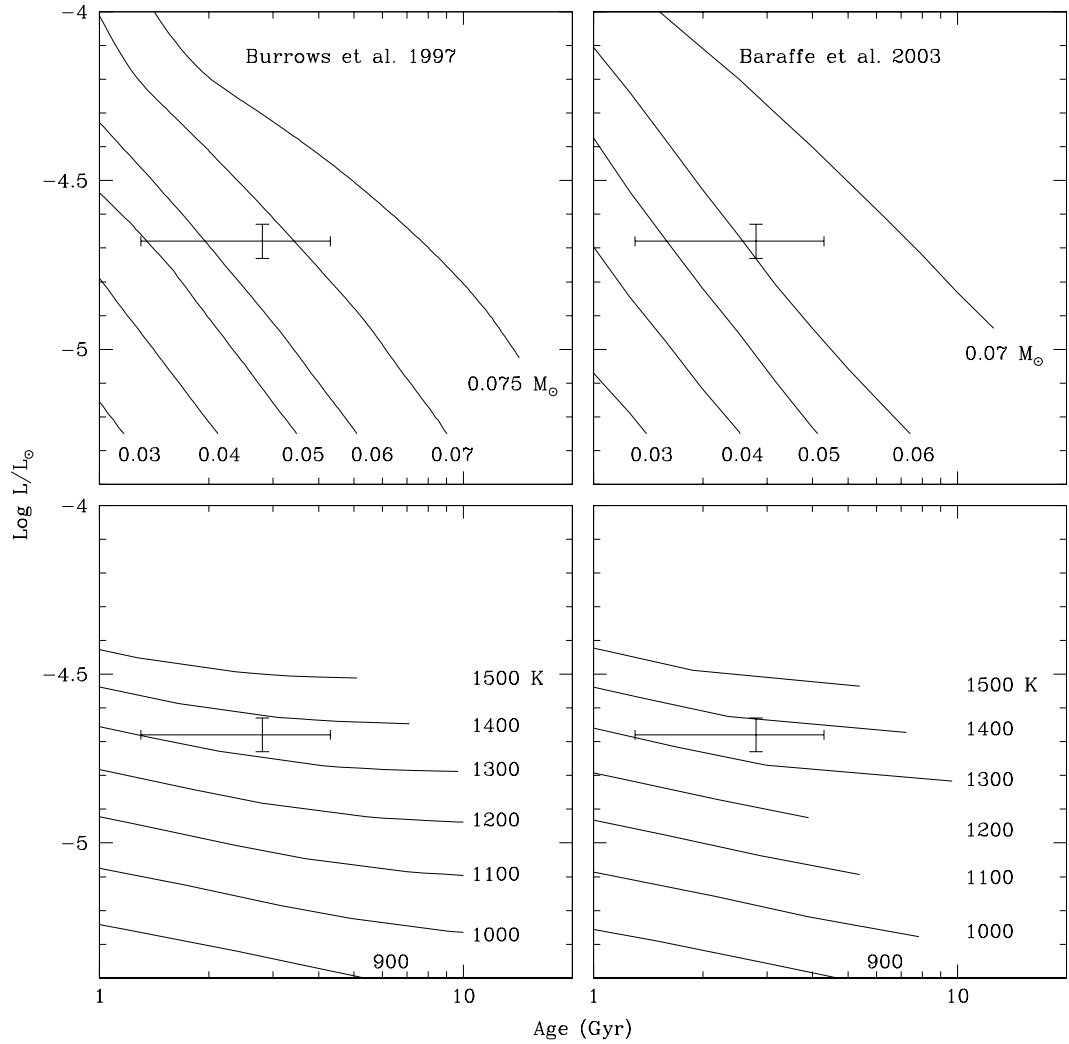}
\caption{
Luminosity of HD~46588~B compared to the
luminosities as a function of age predicted by the theoretical
evolutionary models of \citet{bur97} and \citet{bar03}
for constant values of mass and temperature.
}
\label{fig:models}
\end{figure}

Our proper motion measurements and spectroscopy have demonstrated the
companionship of HD~46588~B. Therefore, it should have the same
distance, age, and metallicity as its primary. In Table~\ref{tab:data},
we list previous measurements and estimates of these parameters for
HD~46588. To estimate the bolometric luminosity of HD~46588~B, we
converted its 2MASS $K_s$ magnitude to the Mauna Kea Observatory (MKO) 
photometric system \citep{ste04} and applied a bolometric correction
of BC$_K=3.1$ \citep{gol04}, the distance of the primary,
and the absolute bolometric magnitude for the Sun ($M_{\rm bol \odot}=4.75$).
We arrive at a value of log~$L/L_\odot=-4.68\pm0.05$.
Based on this luminosity and the age of the primary, we estimate 
masses of $0.067^{+0.005}_{-0.017}$~$M_\odot$ and
$0.061^{+0.005}_{-0.016}$~$M_\odot$
from the evolutionary models of \citet{bur97} and \citet{bar03}, respectively
(see Figure~\ref{fig:models}).
Both of these estimates are fully below the hydrogen burning mass limit
\citep[$\sim0.075$~$M_\odot$,][]{cha00,bur97},
indicating that HD~46588~B is probably a brown dwarf.
To encompass both ranges of masses, we adopt a mass of
$0.064^{+0.008}_{-0.019}$~$M_\odot$.
These models imply an effective temperature of $1360^{+50}_{-80}$~K, which
agrees with previous estimates for late L dwarfs \citep{gol04,vrba04}.
The photometric, astrometric, and physical properties of HD~46588~B
are summarized in Table~\ref{tab:data}. We note that the
WISE Preliminary Release Source Catalog contains an additional photometric
measurement in W3, but the detection is marginal. The companion is not
detected in W4.

\section{Discussion}
\label{sec:disc}

We have presented the discovery of a widely-separated L9 companion
to the nearby F7V star HD~46588.
Among previously known companions, Gl584~C most closely resembles HD~46588~B
in terms of the spectral types of the system components
\citep[G1/G3/L8,][]{kir01}.
Other late-L/early-T companions to stars include Gl337~CD 
\citep[T0,][]{wil01,bur05}, HD~203030~B \citep[L7.5,][]{met06},
HN~Peg~B \citep[T2.5,][]{luh07}, and LHS~2397aB \citep[L7,][]{dup09}.

\begin{figure}
\epsscale{1.0}
\plotone{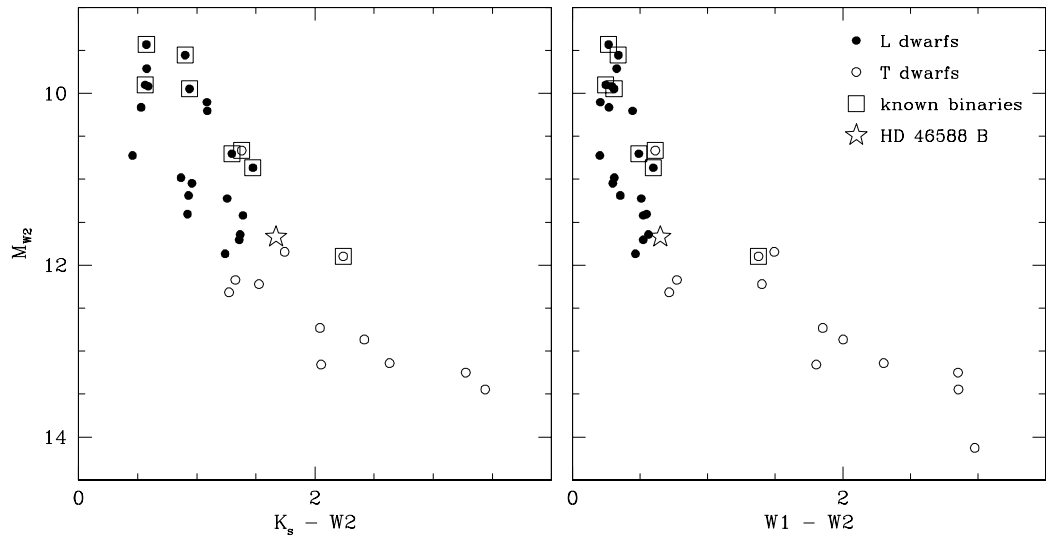}
\caption{
$M_{W2}$ vs.\ $K_s-W2$ and $M_{W2}$ vs.\ $W1-W2$ for HD~46588~B (star) and
L and T dwarfs that have known distances \citep[][references therein]{leg10a}
and that have WISE and $K_s$ (or transformed $K_{MKO}$) measurements available 
(filled and open circles).
}
\label{fig:cmd}
\end{figure}

Brown dwarfs near the L/T transition have served as laboratories for 
studying the complex atmospheric changes that occur between L and T dwarfs
\citep{bur10a}. Observations of binaries have played a central role
in these studies. For instance, systems with spectral types of late
L or early T appear to have a high binary fraction compared to other types
\citep{bur06b,liu06}. This has been explained by rapid evolution of brown
dwarfs through the L/T transition, which causes single transition objects to
be rare and binaries containing earlier and later types to dominate systems
that have composite types near the L/T boundary \citep{bur06b,bur07}.
These binaries often appear overluminous relative to other brown dwarfs
at the same color or spectral type.
HD~46588~B is somewhat overluminous in $M_{W2}$ vs.\ $K_s-W2$ but appears
near the lower envelope of the L/T dwarf sequence in $M_{W2}$ vs.\ $W1-W2$
(Figure~\ref{fig:cmd}). Thus, it does not show clear evidence of binarity.
Regardless of whether it is single or binary, since its distance is known,
HD~46588~B offers an opportunity to better constrain the relations between
spectral type and absolute magnitudes (e.g., $M_{\rm bol}$, $M_K$) near
the L/T transition, which influence the modeling of binary statistics and
surface densities of L/T objects as well as the identification of
overluminous systems \citep{bur07}.
In addition, because of its companionship to a star, HD~46588~B has
the best available age estimate of any known object at a spectral type of L9.
By combining its age and luminosity with evolutionary models, we
have been able to estimate its temperature, and thus constrain the
temperature of the L/T boundary.
The available measurements of $M_{\rm bol}$, $M_K$, and temperature
for HD~46588~B are consistent with those derived previously for companions
with ages of $>$1~Gyr \citep{kir01,dup09}, but these relations could be
more tightly constrained by improving the accuracy of near-IR photometry
for HD~46588~B.

Data from the WISE mission have been recently used to identify new
free-floating late-type members of the solar neighborhood based on their
colors \citep{mai11,bur11,cus11,kir11} and proper motions \citep{sch11,giz11}.
Our discovery of HD~46588~B illustrates the added value of WISE
for uncovering cool companions to nearby stars via their proper motions.
This object also helps to complete the census of L dwarfs within 20~pc
from the Sun \citep{cru07}.

\acknowledgements
We thank Michael Cushing for assistance with reducing the TripleSpec data
and Sandy Leggett for her helpful referee report.
K. L., N. L., and J. B. acknowledge support from grant AST-0544588
from the National Science Foundation.
This publication makes use of data products from the following resources:
the Wide-field Infrared Survey Explorer, which is a joint project of the
University of California, Los Angeles, and the Jet Propulsion
Laboratory/California Institute of Technology, funded by the National
Aeronautics and Space Administration; 
the NASA/IPAC Infrared Science Archive, which is operated by the Jet
Propulsion Laboratory, California Institute of Technology, under contract
with the National Aeronautics and Space Administration; 
the SpeX Prism Spectral Libraries, maintained by Adam Burgasser at
http://pono.ucsd.edu/$\sim$adam/browndwarfs/spexprism;
the M, L, and T dwarf compendium housed at http://DwarfArchives.org and
maintained by Chris Gelino, Davy Kirkpatrick, and Adam Burgasser.
The Center for Exoplanets and Habitable Worlds is supported by the
Pennsylvania State University, the Eberly College of Science, and the
Pennsylvania Space Grant Consortium.

\clearpage

\begin{deluxetable}{lll}
\tabletypesize{\scriptsize}
\tablewidth{0pt}
\tablecaption{Properties of HD~46588~A and B\label{tab:data}}
\tablehead{
\colhead{Parameter} & \colhead{Value} & \colhead{Reference}}
\startdata
\cutinhead{HD~46588 A}
Distance & 17.87$\pm$0.05 pc & 1 \\
$\mu_{\alpha}$ & $-$99.16$\pm$0.19~mas~yr$^{-1}$ & 1 \\
$\mu_{\delta}$ & $-$603.76$\pm$0.26~mas~yr$^{-1}$ & 1 \\
Spectral Type & F7V & 2 \\
Age & 2.8$\pm$1.5~Gyr & 3 \\
$[$Fe/H] & $-$0.21 & 3 \\
\cutinhead{HD~46588 B}
Separation & 79.2$\pm$0.2$\arcsec$ (1420 AU) & 4 \\
Position Angle & 27.46$\pm$0.15$\arcdeg$ & 4 \\
$\mu_{\alpha}$ & $-$112$\pm$19~mas~yr$^{-1}$ & 4,5 \\
$\mu_{\delta}$ & $-$606$\pm$33~mas~yr$^{-1}$ & 4,5 \\
$J$ & 16.26$\pm$0.09 mag & 4 \\
$H$ & 15.08$\pm$0.07 mag & 4 \\
$K_s$ & 14.60$\pm$0.09 mag & 4 \\
W1 & 13.58$\pm$0.03 mag & 5 \\
W2 & 12.93$\pm$0.03 mag & 5 \\
Spectral Type & L9$\pm$1 & 6 \\
log~$L/L_\odot$ & $-$4.68$\pm$0.05 & 6 \\
$T_{\rm eff}$ & $1360^{+50}_{-80}$ K & 6 \\
Mass & $0.064^{+0.008}_{-0.019}$~$M_\odot$ & 6 \\
\enddata
\tablerefs{
(1) \citet{van07};
(2) \citet{abt09};
(3) \citet{cas11};
(4) 2MASS Point Source Catalog;
(5) WISE Preliminary Release Source Catalog;
(6) this work.
}
\end{deluxetable}

\end{document}